\documentclass[preprintnumbers,amsmath,amssymbm,preprint]{revtex4}
\usepackage{epsfig}
\usepackage{graphicx}

\begin{document}
\title{How pure is the tail of gravitational collapse ?}
%\title{High-order terms in the tail of gravitational collapse}
\author{Shahar Hod}
\address{The Ruppin Academic Center, Emeq Hefer 40250, Israel}
\address{ }
\address{The Hadassah Institute, Jerusalem 91010, Israel}
\date{\today}

\begin{abstract}
\ \ \ Waves propagating in a curved spacetime develop tails. In
particular, it is well established that the late-time dynamics of
gravitational collapse is dominated by a power-law decaying tail of
the form $Mt^{-(2l+3)}$, where $M$ is the black-hole mass. It should
be emphasized, however, that in a typical evolution scenario there
is a considerable time window in which the signal is no longer
dominated by the black-hole quasinormal modes, but the leading order
power-law tail has not yet taken over. Higher-order terms may have a
considerable contribution to the signal at these intermediate times.
It is therefore of interest to analyze these higher-order
corrections to the leading-order power-law behavior. We show that
the higher-order contamination terms die off at late times as
$M^2t^{-4}\ln(t/M)$ for spherical perturbations, and as
$M^2t^{-(2l+4)}\ln^2(t/M)$ for non-spherical $(l\neq0)$
perturbations. These results imply that the leading-order power-law
tail becomes ``pure" (namely, with less than $1\%$ contamination)
only at extremely late times of the order of $10^4M$.
\end{abstract}
\bigskip
\maketitle

%]

One of the most remarkable features of wave dynamics in curved
spacetimes is the development of ``tails". Gravitational waves (and
also other fields) propagate not only along light cones, but also
spread inside them. This implies that at late times waves do not cut
off sharply but rather die off in tails. Price \cite{Price} was the
first to analyze the mechanism by which the spacetime outside of a
(nearly spherical) collapsing star divests itself of all radiative
multipole moments, and leaves behind a Schwarzschild black hole.

It is well-established that the late-time dynamics of massless
fields propagating in the curved Schwarzschild spacetime is
dominated by inverse power-law decaying tails, the power indices
equal $2l+3$ \cite{Price}. Physically, these power-law tails are
associated with the backscattering of waves off the effective
curvature potential at asymptotically far regions
\cite{Price,Thorne}.

The analysis of Price has been extended by many authors. Gundlach,
Price, and Pullin \cite{GPP1} showed that power-law tails are a
genuine feature of gravitational collapse-- the existence of these
tails was demonstrated in full non-linear numerical simulations.
Furthermore, since the late-time tail is a direct consequence of the
scattering of the waves at asymptotically far regions, it has been
suggested that power-law tails would develop independently of the
existence of an horizon \cite{GPP2}. It is worth mentioning that the
phenomena of wave tails has also been analyzed in the non-spherical
(rotating) Kerr spacetime
\cite{Hodk1,BarOr,Krivan,Hodk2,Hodk3,Hodk4,Bar,Pullin,Teu}. For
other related works, see
\cite{Bicak,Leaver,Ching1,Ching2,SunPr,Ander,Brad,HodPir1,HodPir2,HodPir3,HodPir4,Bar2,Hodc,
Malec,WangMol,Koy,Mod1,Xue,CarYos,Jing,BerCar,Mod2,SB,He,Gib1,Gib2,Hod1,Hod2}
and references therein.

Despite the flurry of research in this field, our current
understanding of the evolution of wave tails in curved spacetimes is
somewhat unsatisfactory. The leading order inverse power-law tails
are well established, but the resultant formulas are only truly
useful at very late times. In a typical evolution scenario there is
a considerable time window in which the signal is no longer
dominated by the black-hole quasinormal modes, but the leading order
power-law tail has not yet taken over \cite{Ander}. In this
intermediate asymptotic regime (which, as we shall show below, may
be of a very long duration) the signal may have a significant
contribution from higher-order terms. These correction terms
actually contaminate the leading-order power-law tail.

The analysis of higher-order corrections to the leading-order
power-law tail is of theoretical as well as practical importance.
This is especially crucial for the determination of the power index
in numerical simulations which are always characterized by finite
integration times. The precise (pure) power index is expected only
at infinitely late times. However, in any realistic (finite time)
numerical simulation there would be higher-order terms which would
contaminate the pure inverse power-law tails. The influence of these
higher-order terms becomes more significant as the available time of
integration decreases. Thus, in practice, the limited time of
integration introduces an inherent error in the determination of the
power index. It is therefore of great interest to calculate the
higher-order correction terms. The ratio of these contamination
terms to the leading order term would then provide a systematic
indication of the error introduced by the finite-time numerical
simulation. In this work we generalize our previous calculations
\cite{Hodc} to include generic non-spherical perturbations.

We consider the evolution of a massless scalar field in the
spherically symmetric Schwarzschild background (a collapsing star of
a fixed black hole). The external gravitational field is given by
\begin{equation}\label{Eq1}
ds^2=-\Big(1-{{2M}\over r}\Big)dt^2+\Big(1-{{2M}\over
r}\Big)^{-1}dr^2+r^2d\Omega^2\  .
\end{equation}
Resolving the field into spherical harmonics
$\Phi=\sum_{lm}\Psi^{l}_{m}(t,r)Y^{m}_{l}(\theta,\phi)/r$, the wave
equation $\Phi_{;ab}g^{ab}\Phi=0$ for the scalar field in the curved
background reads
\begin{equation}\label{Eq2}
\Big[{{\partial^2}\over{\partial t^2}}-{{\partial^2}\over{\partial
x^2}}+V(x)\Big]\Psi^{l}_{m}=0\  ,
\end{equation}
where the tortoise radial coordinate $x$ is defined by
$dx=dr(1-2M/r)$. The effective curvature potential is given by
\begin{equation}\label{Eq3}
V[r(x)]=\Big(1-{{2M}\over r}\Big)\Big[{{l(l+1)}\over
{r^2}}+{{2M}\over {r^3}}\Big]\  .
\end{equation}
In terms of the tortoise coordinate $x$ and for $x\gg M$ the
scattering potential reads
\begin{eqnarray}\label{Eq4}
V(x)&=&{{l(l+1)}\over{x^2}}+{{4Ml(l+1)\ln
x}\over{x^3}}+{{2M[1-l(l+1)]}\over{x^3}}+{{12M^2l(l+1)\ln^2x}\over{x^4}}\nonumber\\&&+{{4M^2[3-5l(l+1)]\ln
x}\over{x^4}}+O\Big({{M^2}\over{x^4}}\Big)\  .
%{{[12M^2[1-l(l+1)]-8M^2l(l+1)]\ln x}\over{x^4}}
\end{eqnarray}

The dynamics of waves propagating under the influence of logarithmic
scattering potentials of the form
$V(x)=l(l+1)/x^2+\ln^{\beta}x/x^{\alpha}$ (where $\alpha>2$ and
$\beta=0,1$) has been studied in a brilliant work by Ching {\it et
al} \cite{Ching1,Ching2}. The study of wave tails was extended in
Refs. \cite{Hod1,Hod2} to include {\it generic} scattering
potentials. In particular, it has been shown in \cite{Hod1,Hod2}
that the familiar case of logarithmic potentials belongs to a much
wider class of scattering potentials for which
$x{{dV}\over{dx}}/V\to cosnt.$ as $x\to\infty$. For this type of
scattering potentials the late-time tail is dominated by
\cite{Hod1,Hod2}

\begin{equation}\label{Eq5}
\Psi \sim V^{(2l)}(t/2)\  ,
\end{equation}
where $V^{(n)}$ is the $n$th derivative of the function $V(t)$. This
simple result has one exception \cite{Ching1,Ching2,Hod1,Hod2}: if
$\alpha$ is an odd integer less than $2l+3$, the late-time tail is
dominated by
\begin{equation}\label{Eq6}
\Psi \sim V_{ls}^{(2l)}(t/2)\  ,
\end{equation}
where $V_{ls}^{(2l)}$ is the leading subdominant term in the
asymptotic derivative (see \cite{Hod1,Hod2} for details).

We would like to emphasize that the general expressions Eqs.
(\ref{Eq5})-(\ref{Eq6}) are in agreement with, and generalize the
original results of \cite{Ching1,Ching2} which were obtained for the
specific case of logarithmic potentials \cite{Notetech}. In
particular, one finds that for logarithmic scattering potentials the
late-time tail is given by
\begin{equation}\label{Eq7}
\Psi \sim t^{-(2l+\alpha)}\ln^{\beta}t\  .
\end{equation}
This result has the exception that if $\alpha$ is an odd integer
less than $2l+3$, the late-time tail is dominated by
\begin{equation}\label{Eq8}
\Psi \sim t^{-(2l+\alpha)}\ln^{\beta-1}t\  ,
\end{equation}
for $\beta>0$, and by
\begin{equation}\label{Eq9}
\Psi \sim t^{-(2l+2\alpha-2)}\  ,
\end{equation}
for $\beta=0$.

Taking cognizance of the scattering potential Eq. (\ref{Eq4}), one
finds that the late-time tail which characterizes the propagation of
waves in the curved Schwarzschild spacetime is dominated by the
well-known inverse power-law term
\begin{equation}\label{Eq10}
\Psi \sim Mt^{-(2l+3)}\  .
\end{equation}
%\begin{equation}\label{Eq10}
%\Psi \sim -l(l+1)2^{2l+4}{(2l+2)!}Mt^{-(2l+3)}\  ,
%\end{equation}
%for $l>0$, and
%\begin{equation}\label{Eq11}
%\Psi \sim 16Mt^{-3}\  ,
%\end{equation}
%for spherical $l=0$ perturbations.
For the leading subdominant term at asymptotically late times one
finds
\begin{equation}\label{Eq11}
\Psi_{ls} \sim M^2t^{-(2l+4)}\ln^2t\  ,
\end{equation}
for $l>0$, and
\begin{equation}\label{Eq12}
\Psi_{ls} \sim M^2t^{-4}\ln t\  ,
\end{equation}
for spherical $l=0$ perturbations.

We would like to point out that multiple scattering from
asymptotically far regions would also contribute to the late-time
tail. For spherical perturbation ($l=0$) this contribution decays
like $M^2t^{-4}$, and it is therefore negligible as compared to the
leading subdominant term Eq. (\ref{Eq12}) at asymptotically
late-times. On the other hand, for non-spherical perturbations
$(l\neq 0)$ the contribution from multiple scattering goes like
$M^2t^{-(2l+4)}\ln^2t$. This term is of the same order of magnitude
as Eq. (\ref{Eq11}) and it would therefore contribute to the leading
subdominant term of the late-time tail.

In conclusion, we have analyzed the higher-order corrections terms
which ``contaminate" the late-time tail of gravitational collapse.
The existence of a relatively large correction term $\sim
M^2t^{-(2l+4)}\ln^2(t/M)$ appears not to be widely recognized. Aside
from being theoretically interesting, our results are also of
practical importance. They imply that an exact (numerical)
determination of the power index (which characterizes the
leading-order term of the late-time tail) demands extremely long
integration times. (This, in turn, requires the use of highly stable
numerical schemes.) For example, a contamination term of the form
$M^2t^{-(2l+4)}\ln^2(t/M)$ becomes $\sim 1\%$ of the leading-order
term $Mt^{-(2l+3)}$ at extremely late times of the order of $10^4M$.
%For example, for spherical perturbations the leading subdominant
%term (the `contamination') becomes $\sim 1\%$ of the leading-order
%tail at $t\sim 1.1\times 10^4M$. For $l=2$ the term (\ref{Eq12})
%becomes $\sim 1\%$ of the dominant term (\ref{Eq10}) at extremely
%late times of the order of $t\sim 2\times 10^5M$.

\bigskip
\noindent {\bf ACKNOWLEDGMENTS}
\bigskip

This research is supported by the Meltzer Science Foundation. I
thank Tsvi Piran, Liran Shimshi and Yael Oren for stimulated
discussions.

\bigskip

\begin{appendix}\label{App1}
{\bf APPENDIX A: NOTES ON A FORMER WORK}
\bigskip

Following the original work in Ref. \cite{Hodc}, Smith and Burko
(SB) \cite{SB} performed an independent numerical computation of the
late-time tail. These authors failed, however, to find the small
logarithmic correction which accompanies the leading-order power-law
tail. We would like to take this opportunity to make some remarks
about the numerical work of SB.

\begin{itemize}
\item We have emphasized in Ref. \cite{Hodc} that it is essential to develop
an extremely accurate numerical code in order to reveal the
extremely small logarithmic term [which at late-times contaminates
the (already small) leading-order power-law decaying tail]. In
particular, in Ref. \cite{Hodc} we have used a modified version of
the numerical scheme developed in \cite{Gar,HodPiro1,HodPiro2}. This
numerical code is characterized by a {\it fifth}-order convergence
which makes it extremely stable and accurate. On the other hand, the
numerical code used by SB is only of a {\it second}-order
convergence \cite{SB}. The (relatively) poor order-of-convergence of
the code used by SB \cite{SB} may be the reason behind their failure
to reveal the genuine small contamination term.

\item It should be stressed that, actually SB have reported no error
in the analytical derivation of the late-time tail presented in
\cite{Hodc}.

\item Moreover, it is worth emphasizing that SB have {\it not} provided
analytical calculations to support their numerical computations.
(This should be contrasted with the transparent analytical
derivation of the late-time tail presented in Ref. \cite{Hodc}.)

\item We would like to point out the fact that SB provide an
erroneous interpretation of the results of Ching {\it et al}
\cite{Ching1,Ching2}. Ching {\it et al} have shown (see also
\cite{Hod1,Hod2}) that for logarithmic scattering potentials of the
form $V(x)=\ln x/x^{\alpha}$ with $\alpha$ an odd integer less than
$2l+3$, the late-time evolution is characterized by an inverse
power-law decaying tail without a $\ln t$ factor. It should be
stressed, however, that the scattering potential of spherical
($l=0$) perturbations (the ones which were analyzed in
\cite{Hodc,SB}) does not belong to this kind of exceptional
scattering potentials. This clearly shows that the numerical failure
of SB to detect the genuine logarithmic corrections can {\it not} be
explained by the results of Ching {\it et al}. In fact, the results
of Ching {\it et al}, both analytical and numerical
\cite{Ching1,Ching2}, contradict the conclusions of SB.
\end{itemize}

\end{appendix}


\begin{thebibliography}{99}
\bibitem{Notetech} It is worth mentioning that these two studies used different technical
and mathematical tools. The analysis of \cite{Ching1,Ching2} is
based on the spectral decomposition methods of Leaver \cite{Leaver}.
On the other hand, the study of the late-time tail in
\cite{Hod1,Hod2} is based on a modified version of Price's method
\cite{Price}. The outcomes of these two different approaches are in
accord with each other.
\end{thebibliography}
\end{document}